\RequirePackage{fix-cm} 
\documentclass[a4paper,twoside,reqno,12pt]{amsart}

\usepackage{fixltx2e}     
\usepackage{a4wide,verbatim}
\usepackage{amsmath,amsfonts,amssymb,amsgen,amsbsy}
\usepackage{eucal,dsfont,mathrsfs,MnSymbol}

\usepackage{tikz}
\usetikzlibrary{arrows}

\newcommand{\depth}{d}                                 
\newcommand{\sur}[1]{\tilde{#1}}                       
\newcommand{\surs}[1]{\widetilde{#1}}                  
\newcommand{\eqdef}{\stackrel{\text{\tiny{def}}}{=}}   

\newcommand{\ud}{\mathrm{d}}

\newcommand{\ue}{\mathrm{e}}
\newcommand{\ui}{\mathrm{i}}
\newcommand{\upi}{{\pi}}

\newcommand{\half}{{\textstyle{1\over2}}}
\newcommand{\ihalf}{{\textstyle{\ui\over2}}}

\newcommand{\threehalf}{{\textstyle{3\over2}}}

\renewcommand{\Re}{\operatorname{Re}}
\renewcommand{\Im}{\operatorname{Im}}

\newlength{\intwidth}
\DeclareRobustCommand{\fpint}[2]
   {\mathop{%
      \text{%
        \settowidth{\intwidth}{$\int$}%
        \makebox[0pt][l]{\makebox[\intwidth]{$-$}}%
        $\int_{#1}^{#2}$}}}

\newcommand{\plog}[2]{\operatorname{L{\scriptsize{i}}}_{#1}#2}   

\title[Exact relations for steady surface waves]{\bf New exact relations for steady 
irrotational two-dimensional gravity and capillary surface waves}

\author{Didier CLAMOND}

\newcommand{\nfont}{\fontshape{n}\selectfont}
\address{({\nfont\textbf{Didier Clamond}}) Universit\'e C\^ote d'Azur, CNRS-LJAD UMR 7351,
Parc Valrose, F-06108 Nice, France.} 
\email{didierc@unice.fr}

\begin{document}

\maketitle

\begin{abstract}
Steady two-dimensional surface capillary-gravity waves in irrotational motion are considered on 
constant depth. By exploiting the holomorphic properties in the physical plane and introducing some 
transformations of the boundary conditions at the free surface, new exact relations and equations 
for the free surface only are derived. In particular, a physical plane counterpart of the 
Babenko equation is obtained.
\end{abstract}

\section{Introduction}

Although studied for a long time \cite{Craik2004}, steady surface gravity and 
gravity-capillary waves remain a subject of active mathematical research and new 
solutions are still discovered \cite{ClamondEtAl2015,GaoWangVandenB2016}. 
Open questions concern not only extreme gravity waves and `exotic' capillary-gravity  
waves, but also regular gravity waves of finite amplitudes. Indeed, for the latter, many 
of their `obvious' characteristics well-known to physicists remain to be rigorously 
proven mathematically \cite{Constantin2012,Constantin2015,Strauss2010}. In order to 
predict what features are likely true about steady irrotational gravity waves, some numerical 
evidences on the velocity, acceleration and pressure fields were given in \cite{Clamond2012}.   
The present paper is somewhat a continuation of that work, providing new exact relations 
that could be useful for accurate computations, for new mathematical proofs and for 
proving already known results but with weaker assumptions.  

In most mathematical and numerical investigations of two-dimensional surface waves, 
the problem is treated using conformal mapping. This approach, originally due to Stokes 
\cite{Stokes1880}, is appealing because the fluid domain (whose shape is unknown {\em a 
priori\/}) is mapped into a known simple domain (e.g., strip, disk). Thus, in the conformal 
plane, the equations for steady waves can be reduced into a single equation written on a 
known line. Various equations have then been obtained as integral and pseudo-differential 
equations, such as the Babenko \cite{Babenko1987} and Nekrasov \cite{Nekrasov1921} equations 
(see \cite{OkamotoShoji2001,VandenB2010} for reviews). Conformal mapping can also be used 
in presence of vorticity; see \cite{ConstantinEtAl2016} for recent results and conjectures. 

Using conformal mapping is certainly the best approach for computing waves of small or 
finite amplitudes, but it is not so useful for extreme waves. This can be understood, 
for example, considering the limiting gravity wave with a $120^\circ$ inner angle at  
the crest. Such a wave has a power $2/3$ singularity at the crest in the conformal 
plane and, therefore, its $n$-th Fourier coefficient decays like $n^{-5/3}$ as 
$n\to\infty$. Conversely, the same wave in the physical plane has a discontinuous, 
but finite, free surface slope. Therefore, its $n$-th Fourier coefficient decays like 
$n^{-2}$ as $n\to\infty$. (A discussion on the Fourier coefficients of the Stokes waves, 
including the highest one, is given in  \cite{PlotnikovToland2002}.) Another example is 
Stokes' small parameter expansion for waves that are not too steep. It was shown in 
\cite{Clamond2007,DrennanEtAl1992} that these expansions have a better rate of convergence 
in the physical plane than in the conformal plane. The advantage of a physical plane 
formulation was also noticed for an accurate numerical resolution \cite{EvansFord1996} 
and in the derivation of simple accurate analytic approximations \cite{Clamond1999,
Clamond2003,RaineyLH2006}.  
A goal of the present paper is to advocate further the benefit of working in the physical 
plane instead of the conformal one.   

In order to derive relatively simple exact equations in the physical plane, we follow 
the strategy used in \cite{Clamond2013,ClamondCons2013} for the surface reconstruction 
from bottom pressure measurements. In doing so, we obtain equations involving only the free 
surface (i.e., without velocity and pressure fields) that have nice features for analytical 
and numerical investigations. For instance, 
no derivatives need to be computed for gravity waves and only first-order derivatives 
are involved if surface tensions are included.   

The paper is organised as follow. In section \ref{secmathdef}, we introduce the 
notations and precise definitions of the problem. The problem is then reformulated 
in a more tractable form in section \ref{secmathsur}. This reformulation involves 
a holomorphic function $\mathfrak{Q}$ that, for periodic waves, is determined in 
terms of a Fourier expansion in section \ref{seckinfou}. In this section, this 
Fourier series is also summed to obtain a Poisson-like integral formula. The 
definition of the Fourier coefficients, when applied at the free surface, provides 
a system of equations for the free surface, as demonstrated in section \ref{secsyseqeta}. 
In section \ref{secinteq}, integral equations for the free surface are also derived. 
In particular, a physical plane counterpart of the Babenko equation is obtained. 
A regularisation of the integral equation for the free surface is given 
in the section \ref{secregint}, which should be suitable for numerical computations. 
Finally, a summary and some perspectives are given in section \ref{secconclu}.

\section{Definitions and notations} \label{secmathdef}

Steady two-dimensional potential flows due to surface capillary-gravity waves in water 
of constant depth are considered. The fluid is homogeneous of with a constant density 
$\rho>0$. The pressure is denoted $P$ and, at the impermeable free surface, it is equal 
to the surface tension plus the constant atmospheric pressure $P_\text{atm}$; 
$p\eqdef(P-P_\text{atm})/\rho$ denotes the relative scaled pressure. The seabed is fixed, 
horizontal and impermeable.

Let $(x,y)$ be a Cartesian coordinate system moving with the wave, $x$ being the 
horizontal coordinate and $y$ the upward vertical one. The wave is ($2\upi/k$)-periodic 
($k=0$ for aperiodic waves) and $x=0$ is the abscissa of a highest wave crest (subharmonic 
bifurcations yield crests of different heights) or of a deepest trough for aperiodic waves 
of depression when there is only one axis of symmetry. 

$y=-\depth$, $y=\eta(x)$ and $y=0$ denote, respectively, the equations for the bottom, 
for the free surface and for the mean water level. The latter implies that $\left<\eta
\right>=0$ --- $\left<\bullet\right>$ the Eulerian average operator over one period --- 
i.e.
\begin{equation} \label{defmean}
\left<\,\eta\,\right>\ \eqdef\ {k\over2\,\upi}\int_{-{\pi/k}}^{\,{\pi/k}}\eta(x)\ \ud\/x\ =\ 0.
\end{equation}
$a\eqdef\eta(0)$ denotes the wave amplitude (i.e., the mean level to the crest elevation) and 
$b\eqdef-\eta(\upi/k)$ is the trough elevation; hence $a+b$ is the total wave height if the 
surface varies monotonically from crest to trough. A wave steepness $\varepsilon$ is then 
classically defined as $\varepsilon\eqdef k\/(a+b)/2$.

Let be $\phi$, $\psi$, $u$ and $v$ the velocity potential, the stream function, the 
horizontal and vertical velocities, respectively, such that $u=\phi_x=\psi_y$ and 
$v=\phi_y=-\psi_x$. It is convenient to introduce the complex potential 
$f\eqdef\phi+\ui\/\psi$ (with $\ui^2=-1$) and the complex velocity $w\eqdef u-\ui\/v$; 
these are holomorphic functions of $z\eqdef x+\ui\/y$ (i.e., $f=f(z)$ and $w(z)=\ud f/\ud z$). 
The complex conjugate is denoted with a star (e.g., $z^*=x-\ui\/y$), while over-tildes 
denote the quantities written at the surface --- e.g., $\sur{z}(x)=x+\ui\eta(x)$, 
$\sur{\phi}(x)=\phi(x,y\!=\!\eta(x))$. (Note that, e.g., $\sur{u}=\surs{\ \phi_x\,}
\neq\sur{\phi}_x=\sur{u}+\eta_x\sur{v}$.) Because the free surface and bottom are streamlines, 
$\sur{\psi}=\psi(x,\eta)$ and $\psi(x,-\depth)$ are constants.

The dynamic condition can be expressed in term of the Bernoulli equation,
\begin{equation}\label{bernbase}
2\,p\ +\ 2\,g\,y\ +\ u^2\ +\ v^2\ =\ B,
\end{equation}
where $g>0$ is the acceleration due to gravity and $B$ is a Bernoulli constant. 
At the free surface the pressure reduces to the effect of the surface tension, 
i.e., $\sur{p}=-\tau\,\eta_{xx}\,(1+\eta_x^{\,2})^{-3/2}$, $\tau$ being a surface 
tension coefficient divided by the density. Let $-c$ be the mean velocity at the 
bed such that 
\begin{equation} \label{defce}
c\ \eqdef\ -\left<\, u(x,y\!=\!-d)\,\right>.
\end{equation}
Thus, $c$ is the phase velocity of the wave observed in the frame of reference where  
the mean velocity at the bottom is zero, and $c>0$ if the wave travels in the increasing 
$x$-direction. 
A definition of $c$ is not needed for solving the equations in the frame of reference 
moving with the wave, where the flow is steady. A definition of $c$ is however necessary 
when the solution needs to be eventually expressed in another frame of reference. The 
latter can be obtained via a Galilean transformation properly done \cite{Clamond2017a}.

Note that many different phase velocities can be defined (i.e., many frames of 
reference can be of practical interest), and defining phase velocities is not a 
trivial matter, in general \cite{DaSilvaPeregrine1988}. Here, $c$ is Stokes' first definition 
of wave celerity \cite{Clamond2017a,Stokes1847} and comparisons with another phase 
velocity can be found in \cite{Constantin2013}.     
Note also that this frame issue also appears when investigating, for example, fluid particle  
trajectories \cite{Constantin2006,OkamotoShoji2012,Umeyama2012} since the latter are different 
in different frames (i.e., trajectories are not Galilean invariant). In the frame of reference 
moving with the wave, trajectories coincide with streamlines (and streaklines) because the flow 
is steady in this peculiar frame. Trajectories in any other frames can then be obtained via Galilean 
transformations. Note finally that $B=c^2$ in deep water and for solitary waves (but $B\neq c^2$ 
in general) and that $c$ is not the linear phase velocity $c_0\eqdef\sqrt{(g+\tau k^2)
\tanh(k\depth)\//\/k\,}$. 

All the equations in this paper are for non-overturning waves, so 
$\eta$ is a continuous function of $x$. For pure gravity waves, there are no overhanging 
travelling waves of permanent form \cite{Spielvogel1970,Varvaruca2008}. 
The situation is different in presence of surface tensions where overhanging 
solutions exist \cite{LH1989,VandenB2010}. 
Because we derive below some equations for $\eta$ only, their generalisation for 
overturning waves is straightforward, using  for instance the arc length coordinate 
as independent variable. Thus, using $x$ as independent does not really restrict the 
generality of the derivations below.

\section{Surface dynamic relations} \label{secmathsur}

A steady impermeable free surface implying that $\tilde{\psi}$ is a constant, 
the Bernoulli equation (\ref{bernbase}) at the free surface can be written
\begin{equation} \label{berste}
2\,g\,\eta\ +\ \tilde{\phi}_x^{\,2}\,(1+\eta_x^{\,2})^{-1}\ -\ 
2\,\tau\,\eta_{xx}\,(1+\eta_x^{\,2})^{-3/2}\ =\ B.
\end{equation}
Hence, the Bernoulli constant $B$ is such that
\begin{equation} \label{defB}
B\ =\,\left<\,\tilde{\phi}_x^{\,2}\,(1+\eta_x^{\,2})^{-1}\,\right>\,=\,\left<
\phi_x^{\,2}(x,-\depth)\right>,
\end{equation}
the first equality being a consequence of the condition  (\ref{defmean}) and of the periodicity 
applied to (\ref{berste}), while the second equality derives from the irrotationality 
\cite{ClamondCons2013,LH1975}. 

The relation $w=\ud\/f/\ud\/z$ written at the free surface, with the equation 
(\ref{berste}), yields \cite{ClamondCons2013}
\begin{align} \label{defw2}
\tilde{w}^{\,2}\ =&\,\left(\/\frac{\ud\,\tilde{f}}{\ud\/\tilde{z}}\,\right)^{\!2}
\,=\,\left(\frac{\ud\,\tilde{f}}{\ud\/x}\left/\,\frac{\ud\,\tilde{z}}{\ud\/x}\right.\right)^{\!2}
\,=\,\left({\tilde{\phi}_x\over 1+\ui\eta_x}\right)^{\!2}\,=\ \frac{\tilde{\phi}_x^{\,2}}
{1+\eta_x^{\,2}}\,{1-\ui\eta_x\over1+\ui\/\eta_x} \nonumber \\
 =&\,\left[\,B\ -\ 2\,g\,\eta\ +\ 2\,\tau\,\eta_{xx}\,(1+\eta_x^{\,2})^{-3/2}\,\right]
(1-\ui\eta_x)\,/\,(1+\ui\/\eta_x).
\end{align}
$w^2$ being a holomorphic function, the function \cite{Clamond2013}
\begin{equation} \label{defQ}
\mathfrak{Q}(z)\ \eqdef\ \int_{z_0}^z\half\left[\,B\,-\,w^2(z')\,\right]\,\mathrm{d}z',
\end{equation}
($z_0$ an arbitrary constant) is of course also holomorphic. Taking $z_0$ at the wave crest 
(i.e., choosing $z_0=\ui\/a$), integrating along the surface and using (\ref{defw2}), one 
obtains at once
\begin{align} \label{eqQs}
\widetilde{\mathfrak{Q}}(x)\ &=\ \int_{0}^{x}\half\left[\,B\,-\,w^2(z')\,\right]
\left[\,1+\ui\/\eta_x(x')\,\right]\,\mathrm{d}x' \nonumber \\
&=\ g\,H\ -\ \ihalf\,g\left(\eta-a\right)\/
\left(\eta+a-2B/g\right)\ +\ \ui\,\tau\ -\ 
\tau\,(\ui+\eta_x)\,/\,\sqrt{\,1+\eta_x^{\,2}\,},
\end{align}
where $H(x)\eqdef\int_0^x\eta(x')\,\ud\/x'$. For symmetric waves, $H$ is a 
periodic odd function and therefore averages to zero because $\eta$ is a 
periodic even function averaging to zero. For asymmetric waves, $H$ is still 
periodic but, {\em a priori\/}, does not necessarily have zero average.

It follows from (\ref{defB}) and from the bottom impermeability that on the seabed
\begin{equation}\label{eqbotqz}
\left<\,\Re\{\mathfrak{Q}_z\}\,\right>\,=\ 0 \qquad 
\text{and}\qquad \Im\{\mathfrak{Q}_z\}\ =\ 0\ \qquad
\text{at}\quad z\ =\ x\,-\,\ui\/\depth.
\end{equation}
Thus, $\mathfrak{Q}$ is a $(2\pi/k)$-periodic function in the $x$-direction that is bounded 
everywhere at and below the free surface, even in the deep water limit $\depth\to\infty$.

The mathematical formulation involving the function $\mathfrak{Q}$ has the great advantage 
that the conditions at the free surface are combined into a single complex equation. Thus, 
the holomorphic properties can be exploited in a straightforward efficient way, as shown 
in \cite{Clamond2013}.

\section{Fourier expansion and Poisson-like integral} \label{seckinfou}

For a $(2\pi/k)$-periodic wave, such that the boundary conditions (\ref{eqbotqz}) are fulfilled, 
a general solution can be sought as the Fourier expansion
\begin{align} \label{fourierq}
\mathfrak{Q}\ =\ \ui\,\mathfrak{q}_0\ +\ \ui\sum_{n=1}^\infty\left[\,\mathfrak{q}_n\,
\ue^{-\ui\/n\/k\/z}\,-\,\mathfrak{q}_n^*\,\ue^{\ui\/n\/k\/(z+2\/\ui\/\depth)}\,\right], 
\end{align}
where all the Fourier coefficients $\mathfrak{q}_n$ are real if the wave is symmetric with 
respect to the vertical axis $x=0$. The condition $\widetilde{\mathfrak{Q}}(0)=0$ (from the 
definition of $\mathfrak{Q}$) yields
\begin{equation}\label{defq0}
\mathfrak{q}_0\ =\ \sum_{n=1}^\infty\,\left[\,\mathfrak{q}_n^*\,\ue^{-2\/n\/k\/(a+d)}\ 
-\ \mathfrak{q}_n\,\right]\ue^{n\/k\/a}.
\end{equation}
For later convenience, we introduce another holomorphic function $q$:
\begin{align} \label{defq}
q(z)\ \eqdef\ \mathfrak{Q}(z)\ -\ \ui\,\mathfrak{q}_0\ =\ \ui\sum_{n=1}^\infty
\left[\,\mathfrak{q}_n\,\ue^{-\ui\/n\/k\/z}\,-\,\mathfrak{q}_n^*\,\ue^{\ui\/n\/k\/
(z+2\/\ui\/\depth)}\,\right], 
\end{align}
such that at the bottom $\left<q(x-\ui\depth)\right>=0$, $\Im\{q(x-\ui\depth)\}=0$ 
and at the free surface $\left<\sur{q}\/\sur{z}_x\right>=0$.

The coefficients $\mathfrak{q}_n$ can be obtained from $\sur{\mathfrak{Q}}(x)=
\mathfrak{Q}(\sur{z})$ via the relations (spectral projections)
\begin{align} \label{calcqn}
\left<\,\mathfrak{Q}(\sur{z})\,\exp(\ui\/n\/k\/\sur{z})\,\frac{\ud\,\sur{z}}
{\ud\/x}\,\right>\, =\
 \left\{\begin{array}{l}
  \ui\,\mathfrak{q}_n \\
  -\,\ui\,\mathfrak{q}_{-n}^*\,\exp(2\/n\/k\/\depth) \\
  \end{array}\right. \qquad &
  \begin{array}{l}
    \text{if}\quad n\geqslant0, \\
     \text{if}\quad n<0,
  \end{array}
\end{align}
the equality deriving from the obvious relation
\begin{align}
\left<\,\exp(\ui\/n\/k\/\sur{z})\,\frac{\ud\,\sur{z}}{\ud\/x}\,\right>\, =\,
  \left\{\,\begin{array}{l}
    1 \\
    0
  \end{array}\right. \qquad &
  \begin{array}{l}
    (n=0), \\
    (n\neq0).
  \end{array}
\end{align}
Integrating by parts and using the condition (\ref{defmean}), one derives easily the 
similar relations
\begin{align} 
\left<\,\eta\,\exp(\ui\/n\/k\/\sur{z})\,\frac{\ud\,\sur{z}}{\ud\/x}\,\right>\, =&\,
  \left\{\,\begin{array}{l}
    0 \\
    -\left.\left<\,\exp(\ui\/n\/k\/\sur{z})\,\right>\,\right/\/(nk)
  \end{array}\right. &
  \begin{array}{l}
    (n=0), \\
    (n\neq0),
  \end{array} \label{scarel1}\\
\left<\,\eta^2\,\exp(\ui\/n\/k\/\sur{z})\,\frac{\ud\,\sur{z}}{\ud\/x}\,\right>\, =&\,
  \left\{\,\begin{array}{l}
    \left<\,\eta^2\,\right> \\
    -2\left.\left<\,(1\!+\!n\/k\/\eta)\,\exp(\ui\/n\/k\/\sur{z})\,\right>\,\right/\/
    (nk)^2
  \end{array}\right. &
  \begin{array}{l}
    (n=0), \\
    (n\neq0),
  \end{array} \label{scarel2}\\
\left<\,H\,\exp(\ui\/n\/k\/\sur{z})\,\frac{\ud\,\sur{z}}{\ud\/x}\,\right>\, =&\,
  \left\{\,\begin{array}{l}
    \left<\,H\,\right>\ -\ \ui\left<\,\eta^2\,\right> \\
    \ui\left.\left<\,\eta\,\exp(\ui\/n\/k\/\sur{z})\,\right>\,\right/\/(nk)
  \end{array}\right. &
  \begin{array}{l}
    (n=0), \\
    (n\neq0),
  \end{array} \label{scarel3}
\end{align}
as well as (for all $n\in\mathds{Z}$)
\begin{align}
\left<\,\frac{\ui+\eta_x}{\sqrt{1+\eta_x^{\,2}}}\,\exp(\ui\/n\/k\/\sur{z})\,
\frac{\ud\,\sur{z}}{\ud\/x}\,\right>\, =\ \ui\/\left<\,\sqrt{1+\eta_x^{\,2}}\,
\exp(\ui\/n\/k\/\sur{z})\,\right>, \label{scarel4} 
\end{align}
where \(\left<H\right>\neq0\) only (perhaps) for asymmetric waves.
Note that the right-hand sides of (\ref{scarel1}), (\ref{scarel2}) and (\ref{scarel3}) 
do not involve the derivatives of $\eta$, which is an interesting feature for computations.
With the relations (\ref{scarel1})--(\ref{scarel4}), the substitution of (\ref{eqQs}) into 
(\ref{calcqn}) yields
\begin{align}
\mathfrak{q}_0\ =&\ \half\,g\,a^2\ -\ \threehalf\,g\left<\/\eta^2\/\right>\ -\ B\,a\  
-\ \ui\,g\left<\/H\/\right>\,+\ \tau\ -\ \tau\left<\/\sqrt{\/1+\eta_x^2\/}\,\right>  
\qquad &(n=0), \label{solq0} \\
\mathfrak{q}_{n}\ =&\ \left<\left[\,\frac{g}{(n\/k)^2}\,-\,\frac{B}{n\/k}\,+\,
\frac{2\/g\/\eta}{n\/k}\,-\,\tau\/\sqrt{\/1+\eta_x^2\/}\,\right]\/
\ue^{\ui\/n\/k\/\sur{z}}\,\right> \qquad &(n>0), \label{solqn+} \\
-\,\mathfrak{q}_{-n}^*\,\ue^{2\/n\/k\/\depth}\ =&\ \left<\left[\,\frac{g}{(n\/k)^2}\,
-\,\frac{B}{n\/k}\,+\,\frac{2\/g\/\eta}{n\/k}
\,-\,\tau\/\sqrt{\/1+\eta_x^2\/}\,\right]\/\ue^{\ui\/n\/k\/\sur{z}}\,\right> 
\qquad &(n<0), \label{solqn-}
\end{align}
which gives all the $\mathfrak{q}_n$ if $\eta$ is known. 
Hence $\mathfrak{Q}$ and $q$ are completely determined in terms of $\eta$, allowing the 
derivation of equations for $\eta$ only, as shown in the sections \ref{secsyseqeta} and 
\ref{secinteq} below. 
Note that the $\mathfrak{q}_n$ (for $n\neq0$) and $q$ at the free surface, i.e. from 
(\ref{eqQs}) and (\ref{solq0}) 
\begin{equation}\label{defqsur}
\sur{q}\ =\ g\,(H-\left<H\right>)\ -\ \ihalf\,g\left(\eta^2-3\left<\eta^2\right>\right)\ 
+\ \ui\,B\,\eta\ -\ \tau\,(\ui+\eta_x)\left(1+\eta_x^{\,2}\right)^{-1/2}\ +\ 
\ui\,\tau\left<\/\sqrt{\/1+\eta_x^2\/}\,\right>,
\end{equation}
do not depend explicitly on the amplitude $a$.  

Once $\eta$ is known, the function $q$ is obtained from its Fourier expansion 
(\ref{defq}). Substituting the relation (\ref{solqn+}), one derives the Poisson-like 
integral relation  
\begin{align} \label{solq}
q(z)\ =\ &\ui\sum_{n=1}^\infty\left[\,\mathfrak{q}_n\,
\ue^{-\ui\/n\/k\/z}\,-\,\mathfrak{q}_n^*\,\ue^{\ui\/n\/k\/(z+2\/\ui\/\depth)}
\,\right] 
\nonumber \\
=\ &\frac{\ui\/k}{2\/\pi}\int_{-{\pi\over k}}^{\pi\over k}\sum_{n=1}^\infty
\left[\,\frac{g}{(n\/k)^2}\,-\,\frac{B\/-\/2\/g\/\eta^\prime}{n\/k}\,+\,
\tau\/\sqrt{\/1+\eta_x^{\prime\,2}\/}\,\right]\left[\,\ue^{\ui\/n\/k\/(\sur{z}^\prime-z)}\,-\, 
\ue^{-\ui\/n\/k\/(\sur{z}^{*\prime}-z-2\/\ui\/\depth)}\,\right]\ud\/x^\prime \nonumber \\
=\ &\frac{\ui\/k}{2\/\pi}\int_{-{\pi\over k}}^{\pi\over k}
\left[\,\frac{g}{k^2}\,\mathscr{L}_2\,-\,\frac{B\/-\/2\/g\/\eta^\prime}{k}\,
\mathscr{L}_1\,+\,\tau\/\sqrt{\/1+\eta_x^{\prime\,2}\/}\,
\mathscr{L}_0\,\right]\ud\/x^\prime, 
\end{align}
with the kernels (for $\nu=0,1,2$)
\begin{equation}
\mathscr{L}_\nu\ \eqdef\ \plog{\nu}{\!\left(\ue^{\ui\/k\/
(\sur{z}^\prime-z)}\right)}\,-\,\left[\,\plog{\nu}{\!\left(\ue^{\ui\/k\/(\sur{z}^
\prime-z^*+2\/\ui\/\depth)}\right)}\,\right]^*\ =\ \plog{\nu}{\!\left(\ue^{\ui\/k\/
(\sur{z}^\prime-z)}\right)}\,-\ \plog{\nu}{\!\left(\ue^{\ui\/k\/(z-\sur{z}^{*
\prime}+2\/\ui\/\depth)}\right)},
\end{equation}
the primes denoting the dependence with respect to the dummy variable $x'$ (e.g., 
$\eta^\prime=\eta(x^\prime)$, $\sur{z}^\prime=x^\prime+\ui\/\eta(x^\prime)$, etc.) and where
\(
\plog{\nu}{(\theta)} \eqdef \sum_{n=1}^\infty\,{\theta^n / n^\nu}
\)
is the $\nu$-th polylogarithm \cite{NIST}. In particular, $\plog{0}{(\theta)}\eqdef
\theta/(1-\theta)$, $\plog{1}{(\theta)}\eqdef-\log(1-\theta)$ and $\plog{2}{(\theta)}$ 
cannot be expressed with simpler functions\footnote{$\plog{2}{}$ is Spence's dilogarithm, 
which is not to be confused with Euler's dilogarithm $\operatorname{dlog}(\theta)\eqdef
\plog{2}{(1-\theta)}$.} but it can be easily computed \cite{ClamondDilog}. 
$\plog{\nu}{(\theta)}$ for $\nu\!\geqslant\!1$ is a single-valued\footnote{Only the principal 
branch (such that $-\pi<\arg(\theta)\leqslant\pi$) of the complex logarithm is considered 
here. This convention defines the branch cut of $\plog{1}{}$, which is carried out to the 
definition of higher-order complex polylogarithms via the recurrence relation 
$\plog{\nu+1}{(\theta)}=\int_0^\theta{\theta^\prime}^{-1}\plog{\nu}{(\theta^\prime)}\,
\ud\theta^\prime$.} function in the cut plane $\theta\in\mathds{C}\backslash[1,\infty[$, 
i.e., with $\theta=\ue^{\ui\/k\/(\sur{z}^\prime-z)}$ the branch cut is defined by 
$x=x^\prime$ and $y\geqslant\eta^\prime$. 
Therefore, the relation (\ref{solq}) is well defined for all $z$ beneath the free surface. 

Once $q$ is known, the velocity field is obtained from the relation $w^2=B-2\,\ud q/\ud z$.
The pressure and other (e.g. acceleration) fields can then be obtained from $w$ via some 
elementary mathematical derivations.

\section{System of equations for the free surface}\label{secsyseqeta}

Changing $n$ by $-n$ in the complex conjugate of (\ref{solqn-}) and subsequently 
substituting (\ref{solqn+}) into the result in order to eliminate $\mathfrak{q}_n$, 
one obtains after some elementary algebra (for all $n>0$)
\begin{align}
\left<\left[\frac{g}{(nk)^2}+\frac{B-2g\eta}{nk}-\tau\sqrt{1+\eta_x^{\,2}}\right]
\ue^{\ui\/n\/k\/\sur{z}^*}+ \left[\frac{g}{(nk)^2}-\frac{B-2g\eta}{nk}-\tau
\sqrt{1+\eta_x^{\,2}}\right]
\ue^{\ui\/n\/k\/(\sur{z}+2\/\ui\/\depth)}\right>=0.  \label{eqetaqnori}
\end{align}
The (infinite) system of equations (\ref{eqetaqnori}), together with the condition 
(\ref{defmean}), determines completely the surface elevation $\eta$. Therefore, this 
system of equations can be used to compute $\eta$. These relations can be also used 
to check the accuracy of numerical solutions obtained from any formulation of the 
problem, thus providing a much more stringent criterion of convergence and accuracy 
than the few well-known integral relations \cite{LH1975,LH1984,Starr1947} often used 
for this purpose. 

For symmetric gravity waves in deep water --- i.e., when $\eta(-x)=\eta(x)$, $\tau=0$, 
$\depth=\infty$ and $B=c^2$ --- the equation (\ref{eqetaqnori}) is significantly 
simplified as
\begin{align} \label{eqetaqnorideep}
\left<\,\left[\,g\,(\/n\/k\/)^{-1}\ +\ c^2\ -\ 2\,g\,\eta\,\right] 
\exp(\/n\/k\/\eta\/) \cos(\/n\/k\/x\/)\,\right>\,=\ 0 \quad\qquad n=1,2,3,\cdots.  
\end{align}
This simple relation is suitable, in particular, for computing the coefficients 
$\mathfrak{a}_{m,j}$ of the Stokes double expansion
\begin{equation}
\eta\ =\ \sum_{m=1}^\infty\,\sum_{j=0}^\infty\, \epsilon^{m+j}\,
\mathfrak{a}_{m,j}\cos(mkx),
\end{equation}
where $\epsilon$ is a small parameter. With (\ref{eqetaqnorideep}), this calculation is not 
much more demanding in the physical plane than in the conformal one, and it has the advantage 
that the Stokes expansion has a better rate of convergence \cite{Clamond2007}. 

It should be noted that (\ref{eqetaqnorideep}) is somehow a physical plane counterpart of the quadratic 
relations between the Fourier coefficients of the Stokes waves obtained by Longuet-Higgins \cite{LH1978} 
in the conformal plane. The relation (\ref{eqetaqnori}) is the generalisation for finite 
depth and surface tensions, whose conformal plane counterpart has apparently never been derived.

\section{Integral equations for the free surface}\label{secinteq}

Applying (\ref{solq}) at the free surface and substituting (\ref{defqsur}) in 
the left-hand side, one obtains
\begin{gather} 
g\,(H-\left<H\right>)\ -\ \ihalf\,g\left(\eta^2-3\left<\eta^2\right>\right)\ 
+\ \ui\,B\,\eta\ -\ \tau\,(\ui+\eta_x)\left(1+\eta_x^{\,2}\right)^{-1/2}\ +\ 
\ui\,\tau\left<\/\sqrt{\/1+\eta_x^2\/}\,\right> \nonumber\\ 
=\ \frac{\ui\/k}{2\/\pi}\fpint{-{\pi\over k}}{\pi\over k}
\left[\,\frac{g}{k^2}\,\surs{\mathscr{L}_2}\ -\ \frac{B\/-\/2\/g\/\eta^\prime}{k}\,
\surs{\mathscr{L}_1}\ +\ \tau\/\sqrt{\/1+\eta_x^{\prime\,2}\/}\,
\surs{\mathscr{L}_0}\,\right]\ud\/x^\prime, \label{solqs}
\end{gather}
$\fpint{}{}$ denoting a singular integral to be evaluated in the sense of Cauchy 
principal value and where
\begin{equation}
\surs{\mathscr{L}_\nu}\ \eqdef\ \plog{\nu}{\!\left(\ue^{\ui\/k\/
(\sur{z}^\prime-\sur{z})}\right)}\,-\,\left[\,\plog{\nu}{\!\left(\ue^{\ui\/k\/(\sur{z}^
\prime-\sur{z}^*+2\/\ui\/\depth)}\right)}\,\right]^*,
\end{equation}
the kernel $\surs{\mathscr{L}_0}$ being singular (hence the principal value integral), 
$\surs{\mathscr{L}_1}$ being weakly singular (with logarithmic singularity) and 
$\surs{\mathscr{L}_2}$ being regular. Thus, splitting the real and imaginary parts of 
(\ref{solqs}), i.e.,
\begin{gather} 
g\,H\ -\ g\left<H\right>\, -\ \tau\,\eta_x\left(1+\eta_x^{\,2}\right)^{-1/2}  \nonumber\\ 
=\ -\/\frac{k}{2\/\pi}\fpint{-{\pi\over k}}{\pi\over k}
\left[\,\frac{g}{k^2}\,\Im\!\left\{\/\surs{\mathscr{L}_2}\/\right\}\, -\ 
\frac{B\/-\/2\/g\/\eta^\prime}{k}\,\Im\!\left\{\/\surs{\mathscr{L}_1}\/\right\}\ +\ 
\tau\/\sqrt{\/1+\eta_x^{\prime\,2}\/}\,\Im\!\left\{\/\surs{\mathscr{L}_0}\/\right\}
\right]\ud\/x^\prime, \label{solqsRe} \\
B\,\eta\ -\ \half\,g\,\eta^2\ +\ \threehalf\,g\left<\eta^2\right>\,-\ \tau\left(1+
\eta_x^{\,2}\right)^{-1/2}\ +\ \tau\left<\/\sqrt{\/1+\eta_x^2\/}\,\right>  \nonumber\\ 
=\ \frac{k}{2\/\pi}\fpint{-{\pi\over k}}{\pi\over k}
\left[\,\frac{g}{k^2}\,\Re\!\left\{\/\surs{\mathscr{L}_2}\/\right\}\, -\ 
\frac{B\/-\/2\/g\/\eta^\prime}{k}\,\Re\!\left\{\/\surs{\mathscr{L}_1}\/\right\}\ +\ 
\tau\/\sqrt{\/1+\eta_x^{\prime\,2}\/}\,\Re\!\left\{\/\surs{\mathscr{L}_0}\/\right\}
\right]\ud\/x^\prime, \label{solqsIm}
\end{gather}
one obtains two conjugate nonlinear singular integro-differential equations for $\eta$. 
Either equation (\ref{solqsRe}) or equation (\ref{solqsIm}) can be used to compute the 
solution, the other one can be used to check the accuracy of the computed approximation. 

For pure gravity waves ($\tau=0$), the imaginary part (\ref{solqsIm}) of (\ref{solqs}) 
becomes 
\begin{gather} 
2\,B\,\eta\ -\ g\,\eta^2\ +\ 3\,g\left<\eta^2\right>\,=\  
\frac{g}{\upi\,k}\int_{-{\pi\over k}}^{\pi\over k}\Re\!\left\{\,
\plog{2}{\!\left(\ue^{\ui\/k\/(\sur{z}^\prime-\sur{z})}\right)}\,-\ 
\plog{2}{\!\left(\ue^{\ui\/k\/(\sur{z}-\sur{z}^{*\prime}+2\/\ui\/\depth)}\right)}
\,\right\}\ud\/x^\prime \nonumber \\
-\ \frac{1}{\upi}\int_{-{\pi\over k}}^{\pi\over k}(B-2g\eta^\prime)\Re\!\left\{\,
\plog{1}{\!\left(\ue^{\ui\/k\/(\sur{z}^\prime-\sur{z})}\right)}\,-\ \plog{1}
{\!\left(\ue^{\ui\/k\/(\sur{z}-\sur{z}^{*\prime}+2\/\ui\/\depth)}\right)}
\,\right\}\ud\/x^\prime, \label{eqetagrav}
\end{gather}
which is a nonlinear weakly singular purely integral (i.e., not differential) equation 
for $\eta$. Equation (\ref{eqetagrav}) is the physical plane counterpart of the Babenko 
equation \cite{Babenko1987}, the latter being expressed in the conformal plane. 
Thus, we name equation (\ref{eqetagrav}) the ``{\em Eulerian Babenko equation}'' (EBE) 
and the original Babenko equation is called here the ``{\em conformal Babenko equation}'' 
(CBE). The CBE is most often written with pseudo-differential operators that can also be  
expressed as convolution integrals. In doing so, one can easily verify that the CBE involves 
hyper-singular integrals, while the EBE involves only weakly-singular integrals. However, 
the integrals of the EBE are not of convolution type and their kernels involve the unknown 
function $\eta$. 
Except for the highest waves, the CBE can be easily and rapidly solved numerically 
\cite{ClamondDuty2017}. However, the CBE is not so convenient for extreme waves for which 
the EBE is expected to be superior.  

The Nekrasov equation \cite{Nekrasov1921} is sometimes used to compute extreme 
gravity waves \cite{ByattS2001,MurashigeOishi2005}. Like the EBE, Nekrasov's integral equation 
is only weakly singular, but this equation is for the angle the free surface makes with the 
horizontal. This angle being discontinuous at the crest of the limiting waves, the numerical 
resolution of the Nekrasov equation is demanding for near-limiting waves.  
Conversely, the EBE being an equation for the surface itself that is a continuous function, 
even for the limiting gravity waves, it has an attractive feature for extreme waves.    

Equation (\ref{solqsIm}) is an EBE generalised to incorporate surface tensions. The CBE 
with surface tension has long been derived and it was used in mathematical and numerical 
analysis \cite{BuffoniEtAl2000,ClamondEtAl2015}. Equation (\ref{solqsIm}) involves first-order 
derivatives while the CBE with surface tension involves second-order derivatives. Therefore, 
equation (\ref{solqsIm}) should be useful for further mathematical and numerical investigations.

\section{Regularised integral equation}\label{secregint}

When capillarity is included, the singular nature of the EBE is not really problematic 
because the surface tension enforces some regularity of the free surface. However, 
for steep pure gravity waves, such as the limiting waves with a $120^\circ$ inner 
angle at the crest, the singular integral is more problematic due to the low 
regularity of $\eta$, even though the integral is only weakly singular. It is 
therefore of practical interest to regularise the equation. 

The integral equation (\ref{solqsIm}) has a (weak, logarithmic) singularity due to the term 
$\plog{1}{\!\left(\ue^{\ui\/k\/(\sur{z}^\prime-\sur{z})}\right)}$ and has a (single pole) 
singularity due to the term $\plog{0}{\!\left(\ue^{\ui\/k\/(\sur{z}^\prime-\sur{z})}\right)}$. 
Both singularities can be removed as follows.
First, the equation (\ref{solqsIm}) is rewritten as
\begin{gather} 
2\,B\,\eta\ -\ g\,\eta^2\ +\ 3\,g\left<\eta^2\right>\,-\ 2\,\tau\left(1+
\eta_x^{\,2}\right)^{-1/2}\ +\ 2\,\tau\left<\/\sqrt{\/1+\eta_x^2\/}\,\right>\,=\  
\frac{g}{\upi\,k}\int_{-{\pi\over k}}^{\pi\over k}\Re\!\left\{\,\surs{\mathscr{L}_2}
\,\right\}\ud\/x^\prime \nonumber\\   
-\ \frac{2\,g}{\upi}\int_{-{\pi\over k}}^{\pi\over k}(\eta-\eta')
\Re\!\left\{\,\surs{\mathscr{L}_1}\,\right\}\ud\/x^\prime\ 
-\ \frac{B\,-\,2\,g\,\eta}{\upi}\int_{-{\pi\over k}}^{\pi\over k}\Re\!\left\{\,
\surs{\mathscr{L}_1}\,\right\}\ud\/x^\prime \label{eqetagravbis} \\
+\ \frac{k\,\tau}{\pi}\int_{-{\pi\over k}}^{\pi\over k}\left(\sqrt{\/1+\eta_x^{\prime\,2}\/}
-\sqrt{\/1+\eta_x^{\,2}\/}\,\right)\,\Re\!\left\{\,\surs{\mathscr{L}_0}\,\right\}
\ud\/x^\prime\ +\ \frac{k\,\tau}{\pi}\sqrt{\/1+\eta_x^{\,2}\/}\fpint{-{\pi\over k}}{\pi\over k}
\Re\!\left\{\,\surs{\mathscr{L}_0}\,\right\}\ud\/x^\prime, \nonumber
\end{gather}
where only the third and fifth integrands are singular. Second, considering the relations
\begin{gather}
\overline{\mathscr{L}_\nu}\ \eqdef\ \plog{\nu}{\!\left(\ue^{\ui\/k\/(x^\prime-x)}\right)}\,
-\ \plog{\nu}{\!\left(\ue^{\ui\/k\/(x-x^\prime+2\/\ui\/\depth)}\right)}, \qquad
\int_{-{\pi\over k}}^{\pi\over k}\plog{\nu}{\!\left(\ue^{\ui\/k\/(x^\prime-x+\ui\alpha)}\right)}
\,\ud\/x^\prime\ =\ 0, 
\end{gather}
where $\alpha\in\mathds{R}^+$ is a constant, 
the equation (\ref{eqetagravbis}) can be further rewritten
\begin{gather} 
2\,B\,\eta\ -\ g\,\eta^2\ +\ 3\,g\left<\eta^2\right>\,-\ 2\,\tau\left(1+
\eta_x^{\,2}\right)^{-1/2}\ +\ 2\,\tau\left<\/\sqrt{\/1+\eta_x^2\/}\,\right>\,= \nonumber\\   
-\ \frac{2\,g}{\upi}\int_{-{\pi\over k}}^{\pi\over k}(\eta-\eta')
\Re\!\left\{\,\surs{\mathscr{L}_1}\,\right\}\ud\/x^\prime\ 
+\ \frac{k\,\tau}{\pi}\int_{-{\pi\over k}}^{\pi\over k}\left(\sqrt{\/1+\eta_x^{\prime\,2}\/}
-\sqrt{\/1+\eta_x^{\,2}\/}\,\right)\,\Re\!\left\{\,\surs{\mathscr{L}_0}\,\right\}
\ud\/x^\prime\nonumber\\ 
+\ \frac{g}{\upi\,k}\int_{-{\pi\over k}}^{\pi\over k}\Re\!\left\{\,\surs{\mathscr{L}_2}\,-\,
\overline{\mathscr{L}_2}\,\right\}\ud\/x^\prime\ -\ \frac{B\,-\,2\,g\,\eta}{\upi}\int_{-{\pi\over k}}
^{\pi\over k}\Re\!\left\{\,\surs{\mathscr{L}_1}\,-\,\overline{\mathscr{L}_1}\,\right\}\ud\/x^\prime
\nonumber\\
+\ \frac{k\,\tau}{\pi}\sqrt{\/1+\eta_x^{\,2}\/}\int_{-{\pi\over k}}^{\pi\over k}
\Re\!\left\{\,\surs{\mathscr{L}_0}\,-\,(1+\ui\eta_x)^{-1}\,\overline{\mathscr{L}_0}\,\right\}\ud\/x^\prime. 
\label{eqetagravter}
\end{gather}
All the integrands in (\ref{eqetagravter}) are continuous if $\eta$ is continuous (non-overturning 
waves) because as $x^\prime\to x$ we have  
\begin{align}
\Re\!\left\{\,\surs{\mathscr{L}_0}\,-\,(1+\ui\eta_x)^{-1}\,\overline{\mathscr{L}_0}\,\right\}\, 
&\to\ O\!\left(\/(x^\prime-x)^0\/\right), \label{appL0L0}\\
\left(\sqrt{\/1+\eta_x^{\prime\,2}\/}
-\sqrt{\/1+\eta_x^{\,2}\/}\,\right)\,\Re\!\left\{\,\surs{\mathscr{L}_0}\,\right\}\,&\to\ 
\frac{\eta_{xx}\,\eta_x^{\,2}}{k\,(1+\eta_x^{\,2})^{3/2}},\\
\Re\!\left\{\,\surs{\mathscr{L}_1}\,-\,\overline{\mathscr{L}_1}\,\right\}\, 
&\to\ \log\!\left(\/\frac{1-\ue^{-2k(\eta+d)}}{\left(1-\ue^{-2kd}\right)\sqrt{1+
\eta_x^{\,2}}}\/\right),\label{appL1L1} \\
(\eta-\eta')\Re\!\left\{\,\surs{\mathscr{L}_1}\,\right\}\,&\to\ 0,\\
\Re\!\left\{\,\surs{\mathscr{L}_2}\,-\,\overline{\mathscr{L}_2}\,\right\}\, 
&\to\ \plog{2}{\!\left(\ue^{-\/2\/k\/d}\right)}
\ -\ \plog{2}{\!\left(\ue^{-2\/\/k\/(\eta+d)}\right)}, \label{appL2L2} 
\end{align}
the $O\!\left(\/(x^\prime-x)^0\/\right)$ term on the right-hand side of (\ref{appL0L0}) being too 
complicated to be reported here. The additional dilogarithmic term is not necessary because 
$\surs{\mathscr{L}_2}$ is regular as $x^\prime\to x$. The $\overline{\mathscr{L}_2}$ term 
has nevertheless been subtracted in order to improve the equation in view of numerical 
computations. 

When $\tau=0$ (pure gravity waves), (\ref{eqetagravter}) is a regular integral equation for $\eta$.
The absence of derivatives (of the unknown dependent variable $\eta$) in this regular integral equation 
is {\em a priori} an interesting feature for computing extreme waves.

\section{Conclusion}\label{secconclu}  

Several exact relations for steady irrotational surface waves have been derived. 
The derivations were carried out in the physical plane and it was shown that 
these relations are not much more involved than their counterparts in the conformal 
plane. 
In particular, we derived several integral relations for the free surface only 
that can be of practical interest. We also obtained integral equations for the 
free surface that can be used to investigate analytically and numerically steady 
water waves. The main motivation for this work was the derivation of 
equations suitable to study extreme waves, but the practical 
benefit of these equations remains to be demonstrated. Nonetheless, we emphasised 
several of their features that seem advantageous {\em a priori}. It will be the 
subject of future investigations to explore the potential of these equations, for 
instance exploiting equation (\ref{eqetagrav}) to compute accurately the highest 
waves.

The integral relations and equations derived in this paper are valid for non-overturning 
waves, i.e., $\eta$ must be a single valued continuous function of $x$. For 
gravity-capillary waves, some overturning solutions are known to exist \cite{LH1989,VandenB2010}. 
For such solutions, using $x$ as independent variable is no longer possible. One should then 
use instead a parametric representation of the free surface, such as the arc length coordinate. 
The corresponding integral relations and equations can be easily obtained from (\ref{eqetaqnori}), 
(\ref{solqsRe}) and (\ref{solqsIm}). These elementary derivations are left to the reader.

This work could be extended in different directions. Generalisations for overturning waves 
are straightforward using, for instance, the arc length coordinate as independent variable, 
as already mentioned. 
The inclusion of elastic effects at the surface modelling flexural surface waves  
\cite{Toland2008} should also be straightforward. Stratifications 
in homogeneous layers could be treated in similar ways. The inclusion of vorticity would 
be of special interest and it seems conceivable, at least for a constant vorticity. These 
possibilities will be investigated in future works.

\end{document}